\documentclass[prl, twocolumn, superscriptaddress]{revtex4-2}
\usepackage[usenames]{color}
\usepackage[dvipsnames]{xcolor}
\usepackage{etoolbox} 
\usepackage{upgreek}
\usepackage{amsmath}
\usepackage{amssymb} 
\usepackage{graphicx}
\usepackage[utf8]{inputenc}
\usepackage{physics}
\usepackage{empheq}
\usepackage{float}
\usepackage{physics}
\usepackage{siunitx}
\usepackage{soul}
\usepackage{amsfonts}
\usepackage{booktabs}
\usepackage{relsize}
\usepackage{lipsum}
\usepackage{dashbox}
\usepackage{mathtools}
\usepackage{soul}
\usepackage[colorlinks=true,bookmarks=false]{hyperref}
\hypersetup{linkcolor=blue,citecolor=blue,urlcolor=blue} 

\DeclareSIUnit\gauss{G}

\def\cz{\hat{c}_{0,0}}
\def\cp{\hat{c}_{+k,+1}}
\def\cm{\hat{c}_{-k,-1}}
\def\cpm{\hat{c}_{-k,+1}}
\def\cmp{\hat{c}_{+k,-1}}
\def\cx{\hat{c}_{\pm2k_x,0}}
\def\czero{\hat{c}_{0}}

\def\sigmapairs{\sigma(N_\text{p})}
\def\seed{N_\text{p}^\text{s}}
\def\pairs{N_\text{p}}

\newcommand{\beginsupplement}{%
	\setcounter{table}{0}
	\renewcommand{\thetable}{S\arabic{table}}%
	\setcounter{figure}{0}
	\setcounter{section}{0}
	\setcounter{equation}{0}
	\renewcommand{\thefigure}{S\arabic{figure}}
	\renewcommand{\theHfigure}{S\arabic{figure}}
	\renewcommand{\thesection}{S\Roman{section}}
	\renewcommand{\theHsection}{S\Roman{section}}
	\renewcommand{\theequation}{S\arabic{equation}}
	\hypersetup{linkcolor=black,citecolor=blue,urlcolor=blue}  
	
}%

\begin{document}
	
	\title{
		Spin- and Momentum-Correlated Atom Pairs \\ Mediated by Photon Exchange and Seeded by Vacuum Fluctuations
	}
	\author{Fabian Finger}
	\thanks{These authors contributed equally to this work.}
	\author{Rodrigo Rosa-Medina}
	\thanks{These authors contributed equally to this work.}
	\author{Nicola Reiter}
	\author{Panagiotis Christodoulou}
	\author{Tobias Donner}
	\email{donner@phys.ethz.ch}
	\author{Tilman Esslinger}
	\affiliation{Institute for Quantum Electronics and Quantum Center, ETH Zürich, 8093 Zürich, Switzerland}
	\date{\today}
	
	\begin{abstract}
		Engineering pairs of massive particles that are simultaneously correlated in their external and internal degrees of freedom is a major challenge, yet essential for advancing fundamental tests of physics and quantum technologies. In this Letter, we experimentally demonstrate a mechanism for generating pairs of atoms in well-defined spin and momentum modes. This mechanism couples atoms from a degenerate Bose gas via a superradiant photon-exchange process in an optical cavity, producing pairs via a single channel or two discernible channels. The scheme is independent of collisional interactions, fast and tunable. We observe a collectively enhanced production of pairs and probe interspin correlations in momentum space. We characterize the emergent pair statistics and find that the observed dynamics is consistent with being primarily seeded by vacuum fluctuations in the corresponding atomic modes. Together with our observations of coherent many-body oscillations involving well-defined momentum modes, our results offer promising prospects for quantum-enhanced interferometry and quantum simulation experiments using entangled matter waves.
	\end{abstract}
	\maketitle
	
	Correlated pairs of particles have proven pivotal in diverse fields of physics. 
	In condensed-matter systems, strongly correlated phenomena have been interpreted via pairing mechanisms, with the primary example of BCS superconductivity~\cite{Bardeen_1957} where phonon-mediated interactions form electron pairs, correlated simultaneously in their spin and momentum. 
	In cosmology, elementary particle-antiparticle pairs and Hawking radiation emerge out of vacuum fluctuations~\cite{hawking1974,bousso1996}. 
	It is also the vacuum, in quantum-optics experiments, that triggers the production of photon pairs via spontaneous parametric down-conversion~\cite{boyd_nonlinear_2020}, a mechanism of fundamental importance and with applications in quantum technology~\cite{pan_multiphoton_2012}.
	
	Similar mechanisms have been explored with massive particles, paired either in their internal~\cite{chang_coherent_2005,klempt_parametric_2010, lucke_twin_2011, bookjans_strong_2011, gross_atomic_2011, shin_bell_2019, qu_probing_2020} or external~\cite{deng_four-wave_1999, vogels_generation_2002, gemelke_parametric_2005, campbell_parametric_2006, perrin_observation_2007,dall_paired-atom_2009,krachmalnicoff_HeFWM_2010,pertot_collinear_2010,bucker_twin-atom_2011, bonneau_tunable_2013, hodgman_solving_2017,clark_collective_2017,anders_momentum_2021} degrees of freedom. In experiments with quantum degenerate gases, vacuum fluctuations can play an essential role~\cite{leslie_amplification_2009, klempt_parametric_2010,clark_collective_2017,tenart_observation_2021} and facilitate quantum simulation of condensed-matter and cosmological systems~\cite{nation_colloquium_2012, steinhauer_observation_2014,hu_quantum_2019}.
	Metrology with quantum gases, including gravimetry and magnetometry~\cite{pezze_quantum_2018,szigeti_improving_2021,hensel_inertial_2021}, would also benefit from the pairing of massive particles, especially in well-defined spin and momentum modes.
	
	However, typical pairing schemes relying on collisions are limited by the timescales of contact interactions, whereas spurious classical seeds and multimode pair generation limit the achievable metrological advantage~\cite{lewis-swan_sensitivity_2013, guan_tailored_2021}.
	As an alternative, photon-atom pairs can be created at faster timescales in superradiant processes~\cite{inouye_phase-coherent_1999}; yet, comprising different types of particles, they are difficult to manipulate and detect.
	Instead, strong light-matter coupling can be used as a building block to correlate matter pairs in cavity QED systems~\cite{Mivehvar_2021_cavityQED}. 
	This had been demonstrated with single atoms~\cite{hagley1997} and has recently been extended to thermal ensembles, creating pairs from an initial seed~\cite{davis_photon-mediated_2019} and realizing nonlocal~\cite{davis_protecting_2020} and programmable interactions~\cite{periwal_programmable_2021} in the spin degrees of freedom.
	
	Here, we employ a Bose-Einstein condensate (BEC) coupled to a high-finesse optical cavity to generate photon-mediated atom pairs correlated simultaneously in their spin and momentum. 
	Unlike schemes relying on isotropic collisions~\cite{shin_bell_2019,kim_emission_2021}, our implementation directly couples individual momentum modes, offering an efficient route for pair production with large mode occupations in tens of microseconds. 
	The measured pair statistics is consistent with the amplification of vacuum fluctuations in the corresponding atomic modes.
	
	\begin{figure*}[t]
		\centering
		\includegraphics[width=2\columnwidth]{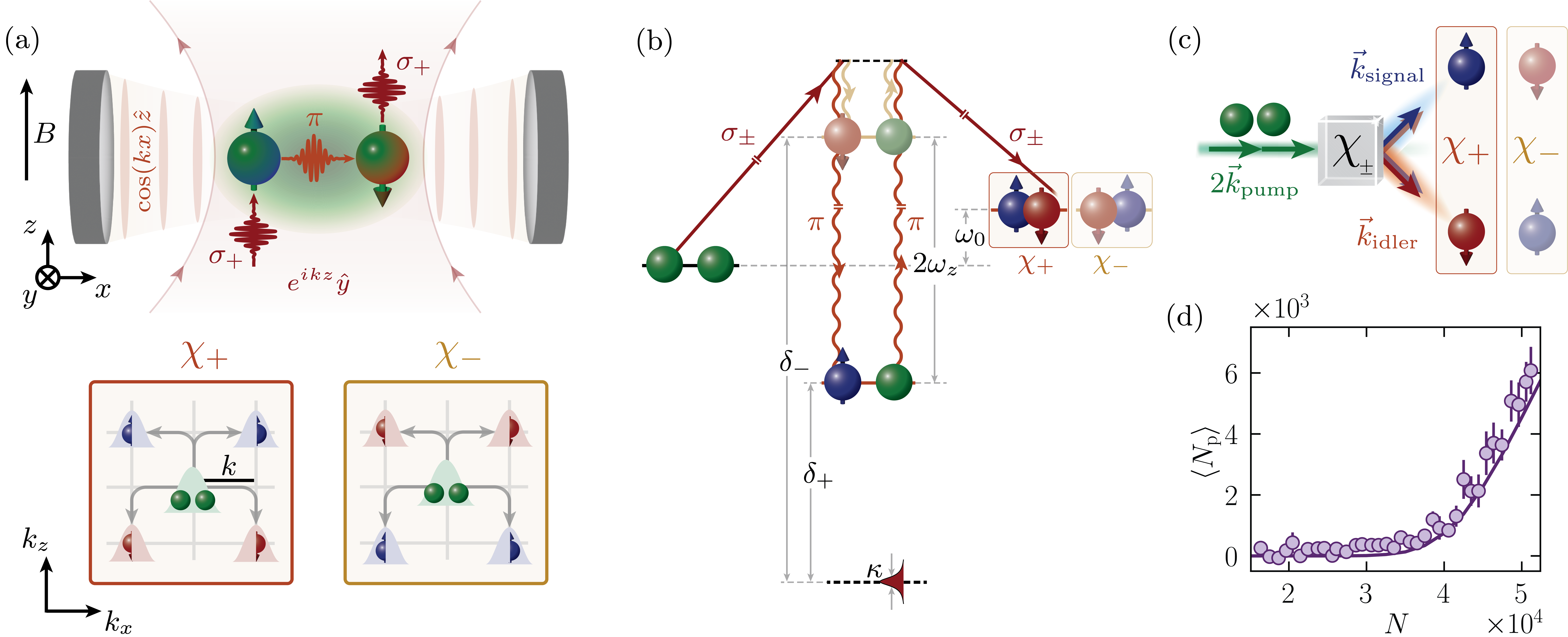}
		\caption{(a)~Microscopic pair-production mechanism. Upper: two $m=0$ atoms in the BEC (green) are converted into a pair of $m=+1$ (blue) and $m=-1$ (red) via two Raman processes involving absorption and emission of a $\sigma_+$ drive photon and a shared virtual cavity $\pi$ photon; arrows on the spheres highlight the acquired spin. With the magnetic quantization axis along $z$, the $y$-polarized running-wave drive is a superposition of $\sigma_\pm$ photons. Thus, pairs can be generated via two channels with rates $\chi_\pm$, as shown in momentum space (lower):  the Raman processes impart recoil momentum $+k$ along the drive ($z$) on the first atom ($m=\pm1$ for $\chi_\pm$ channel) and opposite momentum $-k$ on the second, generating correlated pairs both in spin and in momentum. Because of the standing-wave cavity mode, the pairs acquire also momentum symmetrically along $\pm x$  (illustrated here as half spheres). (b)~Energy-level diagram of the four-photon pairing mechanism, composed of two superradiant Raman processes, each involving a $\sigma_\pm$-polarized drive and a $\pi$-polarized cavity photon, as indicated by straight and curly arrows, respectively. The two intermediate modes are split by twice the linear Zeeman shift $\omega_z$ and correspond to two discernible channels with coupling rates $\chi_\pm$,  depending on the detunings $\delta_\pm$ and the cavity loss rate $\kappa$. The pair energy offset $\omega_0$ is set by the kinetic and internal energy of the pair constituents. (c)~Spin-momentum pair creation as atomic parametric amplifier, with two pump atoms (green) being converted into a pair of signal (blue) and idler (red) atoms via two nonlinear channels $\chi_\pm$. (d)~Measured mean number of pairs $\expval{N_\text{p}}$ after $t=\SI{65}{\micro\second}$, showing a superlinear growth with the initial atom number $N$. Unless specified otherwise, the error bars indicate the standard error. The solid line shows our numerical simulations; see Supplemental Material, also for experimental parameters~\cite{SI}.
		}   
		\label{fig:Fig1}
	\end{figure*}
	
	In our experiments, we prepare a $^{87}$Rb BEC consisting of up to $N\approx8\times 10^4$ atoms in the $m=0$ magnetic sublevel of the $F=1$ hyperfine manifold, with an applied magnetic field $B$ defining the quantization axis $+z$. We couple the atoms dispersively to a single mode of our high-finesse optical cavity~\cite{ottl_hybrid_2006} by illuminating them with a running-wave laser drive propagating along the $z$ direction, cf. Fig.~\ref{fig:Fig1}(a). The drive is switched on for a quench time $t$, and operated at the tune-out wavelength $\lambda=2\pi/k\approx\SI{790.02}{nm}$~\cite{schmidt_precision_2016} to suppress detrimental optical dipole potentials. Overall, this coupling converts atoms in $m=0$ with zero momentum into pairs of $m=\pm 1$ with opposite recoil momenta $\hbar k$ along the drive direction $z$. 
	The typical time required to produce pairs, $T_\text{int}\approx\SI{40}{\micro\second}$, is significantly shorter than time $T_\text{LT}\approx\SI{1}{\milli\second}$ during which the $m=\pm1$ atoms with finite momentum separate from the $m=0$ BEC~\cite{SI}. This separation of timescales, $T_\text{int} \ll T_\text{LT}$, ensures the occupation of well-defined individual momentum modes.
	
	The underlying microscopic mechanism is a superradiant photon-exchange process involving the drive and a single vacuum mode of the cavity field~\cite{davis_photon-mediated_2019}, as illustrated in the upper panel of Fig.~\ref{fig:Fig1}(a). 
	During this process, one atom in the mode $\ket{k_z=0}_{m=0}\equiv\ket{0}_0$ scatters a $\sigma_\pm$ photon from the drive into the $\pi$-polarized cavity mode and flips its spin to $m=\pm 1$, obtaining net recoil momentum $\hbar k$ along $+z$ and occupying the mode $\ket{+k}_{\pm1}$.
	The emitted `virtual' cavity photon is rescattered into the drive field by a second atom in $\ket{0}_0$, which obtains recoil momentum along $-z$ and populates the complementary spin state $m=\mp1$, i.e., the mode $\ket{-k}_{\mp1}$. Crucially, the combination of a BEC and a transverse running-wave drive enables pair production in well-defined spin and momentum modes via two channels depending on the spin flip of the first atom to $m=\pm 1$, cf. lower panel in Fig.~\ref{fig:Fig1}(a). 
	The pairs additionally acquire momentum $\hbar k$ symmetrically in $\pm x$ direction due to the standing-wave structure of the cavity mode. 
	
	To characterize the key properties of our system, we derive an effective many-body Hamiltonian $\hat{H}$ using a few-mode expansion in spin and momentum space, and adiabatically eliminating the cavity field. We obtain $\hat{H}~=~\hat{H}_0 + \hat{H}_+ + \hat{H}_- $, with approximate contributions
	\begin{align}
		\hat{H}_0 &=\frac{\hbar\omega_0}{2}\sum_{\tilde{k}=\pm k}\left(\hat{c}^\dagger_{\tilde{k},1}\hat{c}_{\tilde{k},1}+ \hat{c}^\dagger_{-\tilde{k},-1}\hat{c}_{-\tilde{k},-1} \right),\\
		\hat{H}_{\pm} &= \hbar\chi_\pm \qty(2\hat{c}^\dagger_{\--k,\mp1}\hat{c}^\dagger_{+k,\pm1}\hat{c}_{0,0}\hat{c}_{0,0}+\text{H.c.}),
		\label{Eq2;Hamiltonian}
	\end{align}
	where the bosonic operators $\hat{c}^\dagger_{\tilde{k},m}$ create atoms in the modes $\ket*{\tilde{k}}_m$ with $\tilde{k}=\{0,+k,-k\}$ and $m=\{0,+1,-1\}$~\cite{SI}.
	The various energy scales of the system are schematically depicted in Fig.~\ref{fig:Fig1}(b).
	The first term, $\hat{H}_0$, describes the energy cost $\hbar\omega_0=2 \hbar q + 4\hbar\omega_\text{rec}$ for creating a single pair, with the quadratic Zeeman splitting $q$ and the recoil kinetic energy $\hbar \omega_\text{rec}=h \times 3.68~\text{kHz}$.
	The interaction terms $\hat{H}_{\pm}$ describe the two discernible pair-production channels with the corresponding intermediate states being separated by twice the linear Zeeman splitting $\omega_{z}$.
	The coupling rates $\chi_\pm=\eta^2\delta_\pm/(\delta_\pm^2+\kappa^2)$ depend on the decay rate of the cavity field $\kappa=2\pi\times\SI{1.25}{\mega\hertz}$ and the tunable parameters $\eta$ and $\delta_\pm$, denoting the two-photon scattering rate and detunings of the cavity-mediated Raman processes, respectively~\cite{rosa-medina_observing_2022,SI}.
	The behavior of our system is determined by the competition between $\hat{H}_0$ and $\hat{H}_{\pm}$. For the relevant case of $\delta_\pm<0$, the cavity-mediated interactions are of ferromagnetic character ($\chi_\pm<0$)~\cite{stamper-kurn_spinor_2013}, favoring the formation of pairs in the corresponding atomic modes~\cite{davis_photon-mediated_2019}. 
	
	This pairing mechanism is analogous to spontaneous parametric down-conversion in nonlinear optics, as illustrated in Fig.~\ref{fig:Fig1}(c). The atoms in $\ket{0}_0$ correspond to the input `pump' mode, whereas the finite-momentum atoms in $m=\pm 1$ compare with the output `signal' and `idler' modes. Experimentally, we observe a superlinear increase of the mean pair number $\expval{N_{\text{p}}}$ when adjusting $N$ for a fixed  $t=\SI{65}{\micro\second}$ [Fig.~\ref{fig:Fig1}(d)], in analogy to parametric amplification~\cite{boyd_nonlinear_2020}. This behavior is due to collective enhancement of the pair production, which results in effective coupling rates $N\chi_\pm$ akin to the susceptibility in nonlinear optical media~\cite{boyd_nonlinear_2020}. 
	
\begin{figure}[t]
		\centering
		\includegraphics[width=\columnwidth]{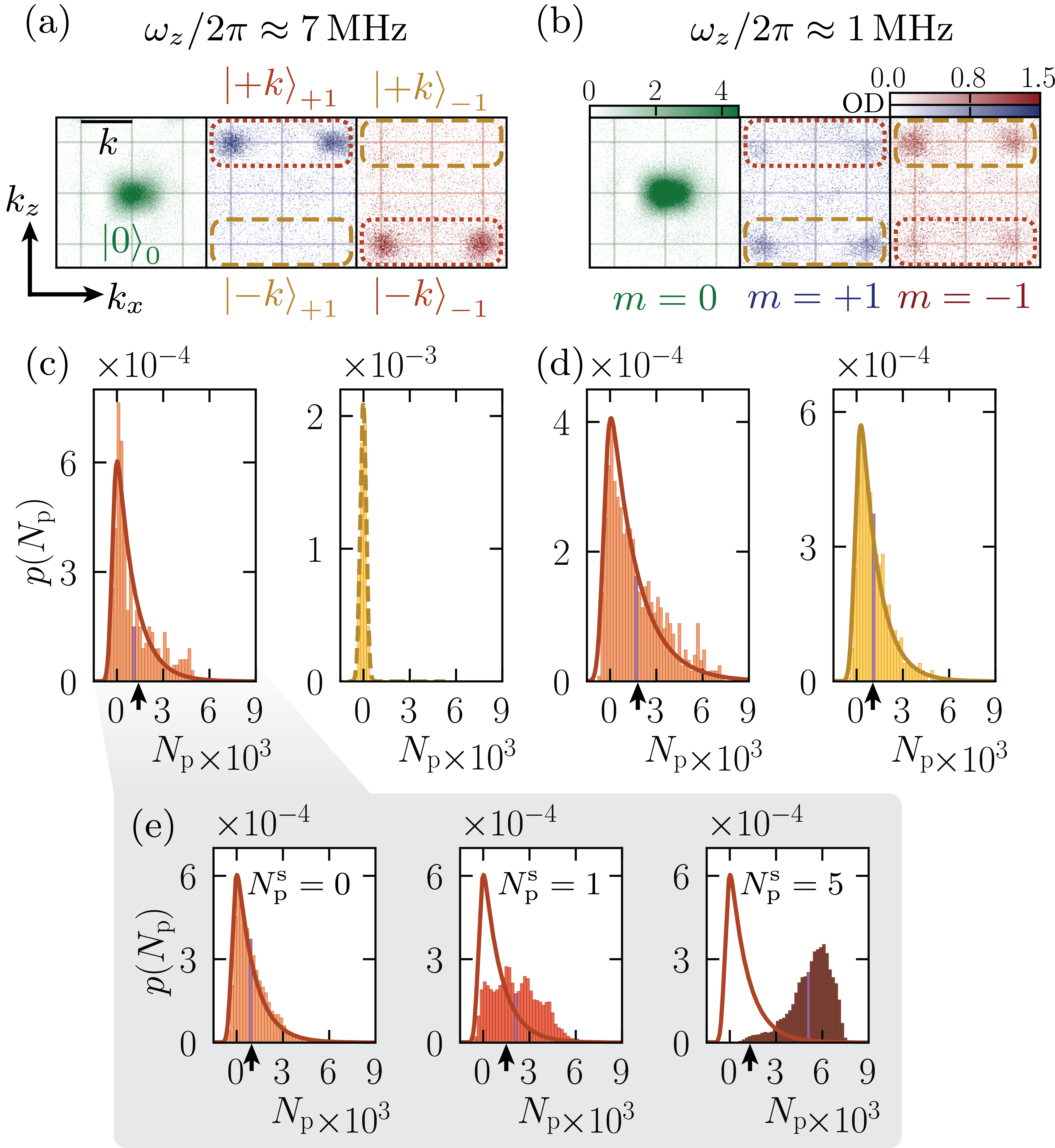}
		\caption{Pair statistics seeded by vacuum fluctuations. (a,~b)~Exemplary spin-resolved momentum distributions for the (a) single-channel and (b) two-channel configuration. The orange and yellow boxes indicate the modes $\ket{\pm k}_{\pm1}$ and $\ket{\pm k}_{\mp1}$, respectively.  (c,~d)~ Pair statistics, generated through the $\chi_+$ (orange) and $\chi_-$ (yellow histograms) process, for (c) the single-channel and (d) two-channel configurations. The solid lines correspond to fitted Bose-Einstein distributions with experimentally determined  mean $\expval{N_\text{p}}$ (purple bin) convolved with our detection noise~\cite{SI}; the dashed line in (c) is consistent with a distribution with zero mean pairs and Gaussian detection noise of $\sim 200$  pairs. The arrows indicate the standard deviation of the resulting distributions, demonstrating $\expval{N_\text{p}} \approx \sigma(N_\text{p})$. (e) Simulated pair statistics convolved with detection noise~\cite{SI} for a single channel and different classical seeds $N_\text{p}^\text{s}$ in the pair modes. The solid line corresponds to the fitted data of (c), and is compatible with the pair distribution being seeded by vacuum fluctuations ($N_\text{p}^\text{s}=0$).} 
		\label{fig:Fig2}
	\end{figure}

	In the experiment, we individually control the coupling rates by varying $\delta_\pm$ via the combined tuning of $\omega_z$ and the cavity resonance and determine the populations of the different atomic modes by measuring spin-resolved momentum distributions~\cite{SI}.  If $\omega_z$ is sufficiently large, only the $\chi_+$ channel contributes and gives rise to pairs occupying the modes $\ket{+k}_{+1}$ and $\ket{-k}_{-1}$. This is highlighted in the exemplary momentum distribution in Fig.~\ref{fig:Fig2}(a) for $\omega_z=2\pi\times\SI{7.09(1)}{MHz}$. For smaller $\omega_z$, both channels become active, resulting in additional occupation of the modes $\ket{+k}_{-1}$ and $\ket{-k}_{+1}$, as shown in Fig.~\ref{fig:Fig2}(b) for $\omega_z=2\pi\times\SI{1.01(1)}{MHz}$. In the following, we will refer to these two settings as the single- and two-channel configurations. 
	To characterize the resulting quantum states, we accumulate hundreds of experimental realizations for the single-channel [Fig.~\ref{fig:Fig2}(c)] and two-channel [Fig.~\ref{fig:Fig2}(d)] configurations and obtain the respective pair statistics. The number of pairs associated with the $\chi_+$ and $\chi_-$ channel are shown in the left and right panels, respectively. When a channel becomes active, we observe large pair-number fluctuations compatible with a Bose-Einstein distribution $p_\text{BE}~= \expval{N_\text{p}}^{N_\text{p}}/(1+\expval{N_\text{p}})^{N_\text{p}+1}$ \cite{perrier_thermal_2019, qu_probing_2020}; this distribution satisfies $\sigma(N_\text{p})\approx \expval{N_\text{p}}$ for the standard deviation $\sigma(N_\text{p})$ and the mean $\expval{N_\text{p}}$, as indicated by the arrow and purple bin in Figs.~\ref{fig:Fig2}(c,d).
	These observations align with our expectation of the system occupying a two-mode squeezed vacuum state, i.e., a superposition of twin-Fock states for the modes $\ket{+k}_{+1}$ and $\ket{-k}_{-1}$ following Bose-Einstein statistics~\cite{SI, yurke_obtainment_1987, qu_probing_2020}. For initially empty signal and idler modes, this state arises from parametric amplification of vacuum fluctuations, analogous to spontaneous parametric down-conversion~\cite{boyd_nonlinear_2020}. 
	
	To further investigate the role of vacuum fluctuations, we compare the single-channel observations in Fig.~\ref{fig:Fig2}(a) to pair distributions obtained from truncated Wigner simulations~\cite{SI}; our simulations stochastically sample quantum fluctuations of the pair modes~\cite{blakie_dynamics_2008} on top of a vacuum state ($N_\text{p}^\text{s}=0$) or classically seeded pair modes ($N_\text{p}^\text{s}>0$), see Fig.~\ref{fig:Fig2}(e). We observe quantitative agreement between the experimental results and the histogram corresponding to only vacuum fluctuations seeding the process, while the classical seeds yield qualitatively distinct distributions. As also studied in BECs undergoing spin-mixing dynamics~\cite{klempt_parametric_2010,lewis-swan_sensitivity_2013}, already small classical seeds [$\mathcal{O}(1)$] would substantially accelerate the pair dynamics, yielding values $\expval{N_\text{p}}$ that greatly exceed $\sigma(N_\text{p})$ [see arrow and purple bin in Figs.~\ref{fig:Fig2}(e)]. The resulting pair statistics thus serves as a sensitive probe for vacuum fluctuations even in the presence of detection noise~\cite{guan_tailored_2021}. 
	Our findings are compatible with the negligible thermal occupation of the pair modes in our system, which we estimate to be $\expval{N_\text{T}}\lesssim 0.016$~\cite{SI}. As shown in Fig.~\ref{fig:Fig2}(d), parametric amplification with discernible channels is not expected to alter the resulting distributions for an undepleted pump mode~\cite{paleri_photonstatistics_2004, lewis-swan_proposal_2015}. We then expect a product state of two-mode squeezed vacuum states for the two discernible channels~\cite{lewis-swan_proposal_2015, SI}, akin to selecting the overlapping modes at the intersection of both polarization cones in spontaneous parametric down-conversion~\cite{pan_multiphoton_2012, white_nonmaximally_1999}. For a detailed comparison, see the Supplemental Material~\cite{SI}.
	
\begin{figure}[t]
		\centering
		\includegraphics[width=\columnwidth]{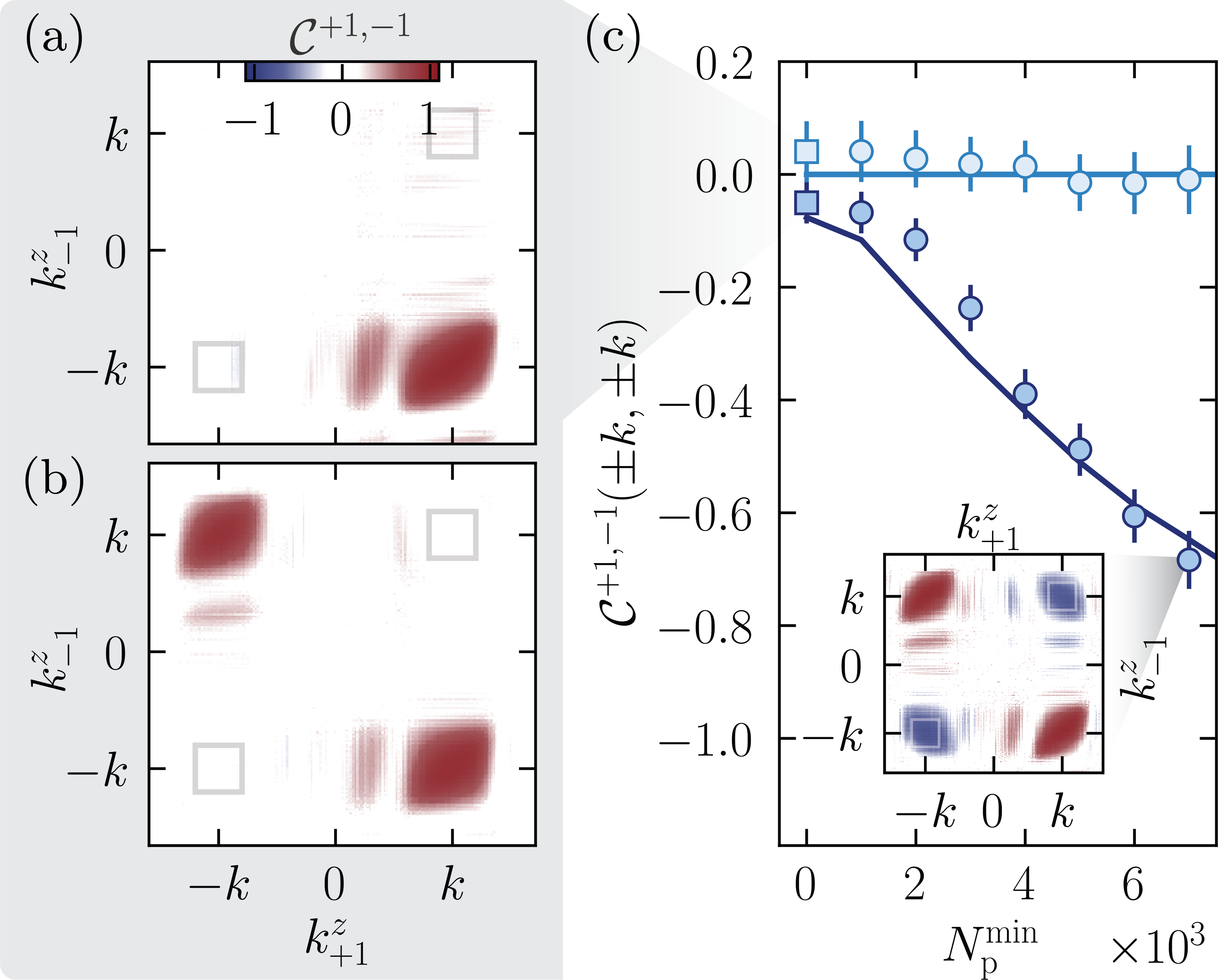}
		\caption{Correlated formation of atomic pairs. (a,~b)~Momentum-space interspin correlation maps $\mathcal{C}^{+1,-1}(k_{+1}^z,k_{-1}^z)$ for the (a) single-channel and (b) two-channel configuration, demonstrating the correlated nature of the produced pairs. We attribute the side patterns close to the correlation peaks to residual density-dependent imaging artifacts. (c)~Anticorrelations $\mathcal{C}^{+1,-1}(\pm k, \pm k)$ for realizations with $N_\text{p} > N_\text{p}^\text{min}$ for a single channel (light blue) and two channels (dark blue), with the solid lines showing the results from our numerical simulations. The anticorrelations are obtained by averaging a region around $\mathcal{C}^{+1,-1}(\pm k, \pm k)$ on the order of the mode sizes~\cite{SI} as schematically shown for the first data points (squares) corresponding to (a) and (b) by the gray squares. The two-channel anticorrelations increase with $N_\text{p}^\text{min}$ due to pump-mode depletion. The inset displays a representative correlation map for $N_\text{p}^\text{min}=7\times 10^3$. The error bars indicate the standard deviation of the averaged region.}
		\label{fig:Fig_new}
	\end{figure}

	Going beyond studies of individual modes, we verify the correlated emission of atom pairs. We introduce the interspin noise correlation map 
	\begin{align}
		\mathcal{C}^{+1,-1}(k_{+1}^z,k_{-1}^z)=&\frac{\expval{n_{+1}n_{-1}} - \expval{n_{+1}}\expval{n_{-1}}}{\sigma({n_{+1}})\sigma({n_{-1}})},
	\end{align}
	with $n_m \equiv n_m(k_{m}^z)$ indicating the momentum-space density distribution of spin state $m$ along $z$ (after integrating along $x$) at coordinate $k_{m}^z$, and $\sigma(n_{m})=\expval{n_{m}^2}-\expval{n_{m}}^2$. In Figs.~\ref{fig:Fig_new}(a) and (b), we show the extracted correlation maps $\mathcal{C}^{+1,-1}(k_{+1}^z,k_{-1}^z)$. For the single-channel configuration, we observe positive correlations around $(k^z_{+1},k^z_{-1})=(+k, -k)$, demonstrating that pairs occupy the modes $\ket{+k}_{+1}$ and $\ket{-k}_{-1}$ in a correlated fashion.
	For two channels, the positive peaks around $(+k, -k)$ and $(-k, +k)$ indicate correlated generation of $m=\pm1$ pairs via both channels, a prerequisite for generating bipartite spin entanglement~\cite{kunkel_spatially_2018,fadel_spatial_2018,lange_entanglement_2018}. When postselecting for realizations above a minimum pair number $N_\text{p}^\text{min}$, we observe increasingly pronounced anticorrelation peaks for the two-channel configuration around equal momenta $(+k,+k)$ and $(-k,-k)$, cf. Fig.~\ref{fig:Fig_new}(c). We attribute this behavior to the competition between the channels in the presence of pump-mode depletion, which inhibits large simultaneous occupation of all pair modes. This suggests, that for large occupations, the many-body state can no longer be expressed as a product state of two-mode squeezed vacuum states for each channel~\cite{SI}.
	
\begin{figure}[t]
		\centering
		\includegraphics[width=\columnwidth]{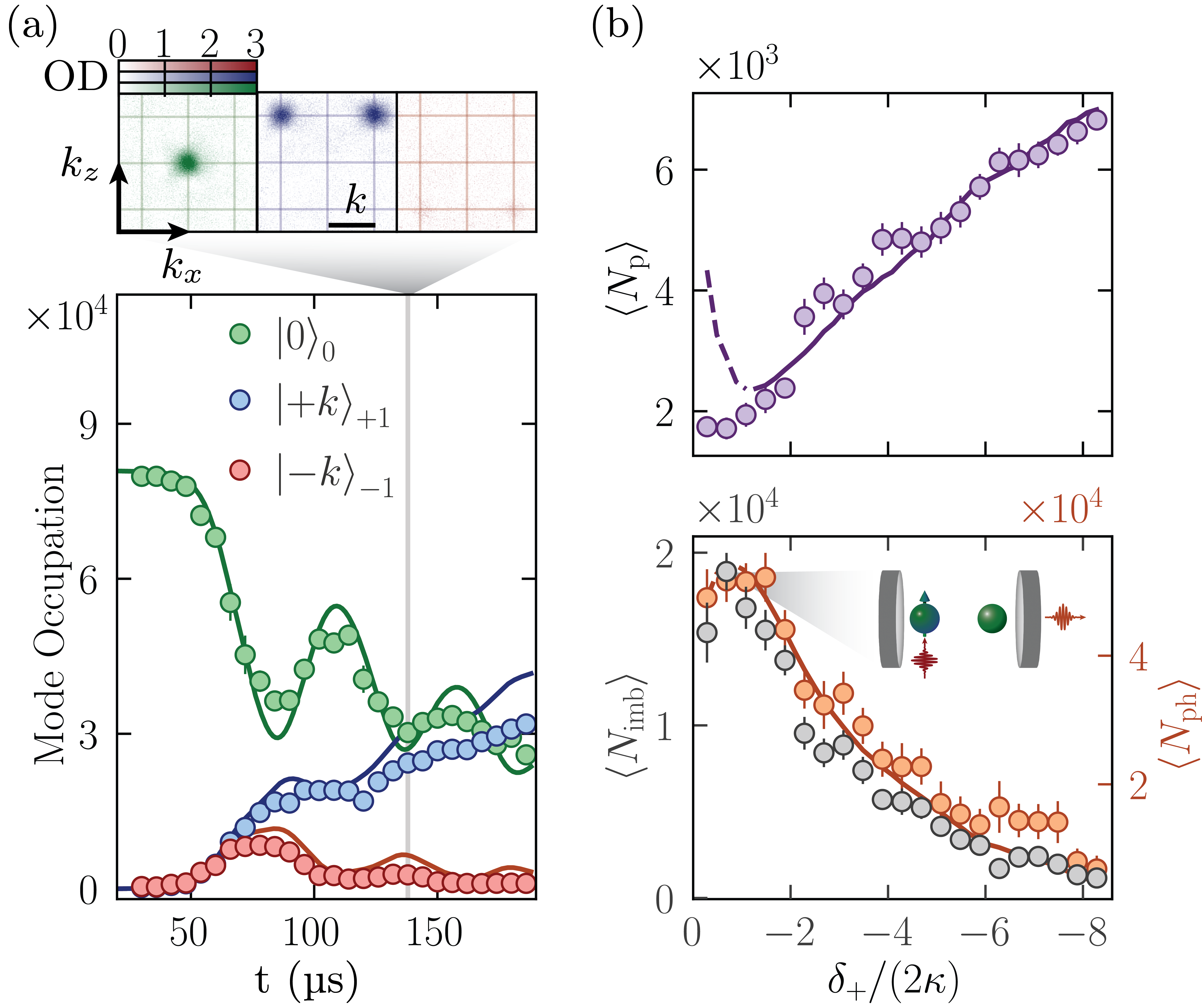}
		\caption{Coherent many-body oscillations and tunable dissipation. (a)~Time evolution of mode occupations in the single-channel configuration [$\delta_+/(2\kappa)=\SI{-9.08(4)}{}$], exhibiting oscillatory dynamics. Photon loss results in a progressive imbalance between the $\ket{+k}_{+1}$ and $\ket{-k}_{-1}$ populations. Inset: representative momentum-space distribution at $t=\SI{138}{\micro\second}$. (b)~Mean number of pairs $\expval{N_\text{p}}$ (upper) and imbalance $\expval{N_\text{imb}}$ (lower,~gray), and number of photons lost from the cavity $\expval{N_\text{ph}}$ (lower,~orange)~\cite{rosa-medina_observing_2022} for $t=~\SI{80}{\micro\second}$ and $\chi_{+,\text{fix}}= -2\pi\times\SI{0.50(2)}{Hz}$ as a function of the detuning $\delta_+/(2\kappa)$, which controls the coherent and dissipative processes. The solid curves show numerical simulations~\cite{SI}, with the dashed line indicating the regime where the adiabatic cavity field elimination becomes invalid, i.e., $\abs*{\delta_+}\lesssim 2\kappa$. We attribute both the deviation from theory at large quench times as well as the excess photon numbers ($\expval{N_\text{ph}}>\expval{N_\text{imb}}$) to superradiant decay to higher-order momentum modes in $m=+1$, which are outside the field of view ($\sim 2.2k$) along $x$ ~\cite{rosa-medina_observing_2022}. Inset: illustration of superradiant scattering.}
		\label{fig:Fig3}
	\end{figure}

	A deeper understanding of the pair dynamics and its interplay with depletion effects can be gained by investigating the full population evolution of the different modes [Fig.~\ref{fig:Fig3}(a)]. For clarity, we concentrate on the single-channel configuration involving the modes $\ket{+k}_{+1}$ and $\ket{-k}_{-1}$. We observe pair production to set in around $T_\text{int}\approx\SI{40}{\micro\second}$, followed by a fast superlinear population increase, in resemblance to optical parametric amplification~\cite{boyd_nonlinear_2020}. As time elapses, we observe coherent many-body oscillations redistributing the atoms between the different available modes, corresponding to the system oscillating around its new ground state with a finite number of pairs~\cite{chang_coherent_2005, stamper-kurn_spinor_2013}.
	While similar to pair oscillations arising from spin-mixing interactions, our observations demonstrate coherent pair dynamics involving well-defined momentum modes. For longer times, we observe a progressive accumulation of atoms in $\ket{+k}_{+1}$ [see inset in Fig.~\ref{fig:Fig3}(a)], resulting in a population imbalance between $\ket{+k}_{+1}$ and $\ket{-k}_{-1}$. The oscillations are damped on a timescale $T_\text{coh}\sim\SI{150}{\micro\second}$, which we identify as the coherence time. 
	
	We attribute both effects to the residual openness of our system as photons are sporadically lost at the cavity mirrors, inhibiting the reabsorption of cavity photons and thereby the formation of the second pair constituent [cf. inset in Fig.~\ref{fig:Fig3}(b)]. We model this dissipative superradiant Raman process via effective Lindblad terms with rates $\gamma_\pm=\eta^2\frac{2\kappa}{\delta_{\pm}^2+\kappa^2}$ for the two channels~\cite{SI}. Our truncated Wigner simulations quantitatively reproduce the observed population evolution [solid lines in Fig.~\ref{fig:Fig3}(a)], with the coupling $\eta$ being the only free parameter of the simulations and optimized to $\eta=0.94\eta_{\text{exp}}$ of the experimentally calibrated value $\eta_{\text{exp}}$. We attribute this small difference to the imperfect alignment between the BEC and the cavity mode, and systematic uncertainties in the atom-number calibration.
	The fast timescales separate pair production from typical dissipation channels in atom systems, such as three-body losses and trapping effects~\cite{weber_2003_threebody,grimm_2000_optical}.
	Our simulations also indicate that for our experimental parameters on average $\sim\chi_+/\gamma_+\approx 10$ pairs are created before the first photon is lost from the cavity. 
	Finally, the scaling of the couplings $\chi_{\pm}/\gamma_{\pm} = \delta_{\pm}/(2\kappa)$ allows us to independently tune the coherent and dissipative processes of our system.
	We demonstrate this control for the single-channel configuration by quenching to a fixed coupling $\chi_{+,\text{fix}}= -2\pi\times\SI{0.50(2)}{Hz}$ and varying the detuning $\delta_+$ at a constant $t=\SI{80}{\micro\second}$, see Fig.~\ref{fig:Fig3}(b). Experimentally, we increase $\eta$ at larger values of  $\abs{\delta_{+}}$, modifying only the dissipative coupling $\gamma_+=2\chi_{+,\text{fix}}\kappa/\delta_+$ for otherwise identical experimental parameters. The measured mean number of pairs $\expval{N_\text{p}}$ (upper panel) remains small close to the two-photon resonance ($\delta_+=0$), and monotonically increases for large detunings $\abs{\delta_+}/(2\kappa)\gg 1$. Concurrently, the mean population imbalance $\expval{N_{\text{imb}}}$ between $\ket{+k}_{+1}$ and $\ket{-k}_{-1}$  (lower panel) exhibits the opposite trend and gradually decreases toward zero for large detunings. We also present the number of photons $\expval{N_\text{ph}}$ lost from the cavity, as measured with our heterodyne detection~\cite{SI}. 
	The qualitative agreement between $\expval{N_\text{ph}}$ and $\expval{N_{\text{imb}}}$ verifies that photon loss is indeed the primary dissipation source. The experimental results are reasonably captured by our numerical simulations. The deviation of the simulated pair number $\expval{N_\text{p}}$ at $\abs{\delta_+}/(2\kappa) \lesssim 1$ is ascribed to the limited validity of the adiabatic elimination of the cavity field in this regime. 
	
	Correlated pairs via two channels open new perspectives for bipartite entanglement in separated atomic clouds~\cite{salvi_squeezing_2018, wasak_bell_2018, lange_entanglement_2018, fadel_spatial_2018, kunkel_spatially_2018, colciaghi_einstein-podolsky-rosen_2023} and loophole-free Bell tests with massive particles~\cite{lewis-swan_proposal_2015,kitzinger_bell_2021} when combined with mode-selective spin rotations. Such nonlocal measurements are particularly fragile against classical seeds~\cite{lewis-swan_sensitivity_2013,evrard_coherent_2021}, highlighting the importance of amplified vacuum fluctuations. Independent control over the sign and strength of the photon-mediated interactions present prospects for implementing time-reversal protocols for noise-resilient atom interferometry~\cite{linnemann_quantum-enhanced_2016, colombo_time-reversal-based_2022,kresic_interferometry_2023}. Finally, extending our scheme to degenerate Fermi gases could facilitate the manipulation of photon-induced Cooper pairs~\cite{colella_quantum_2018,lewis-swan_cavity-qed_2021}.
	
	\begin{acknowledgments}
		We are grateful to Alexander Frank for his contributions to the electronic setup of the experiment and to Florentin Reiter, Davide Dreon, and Meng-Zi Huang for fruitful discussions. We thank Jean-Philippe Brantut for carefully reading the manuscript. We acknowledge funding from the Swiss National Science Foundation: projects No. 200020\_212168 and No. 20QT-1\_205584 and National Centre of Competence in Research Quantum Science and Technology (Grant No. 51NF40-185902), from the Swiss State Secretariat for Education, Research and Innovation
		(SERI), and from EU Horizon2020: European Research Council advanced grant TransQ (projects No. 742579).
	\end{acknowledgments}
	
	%

	
	\beginsupplement
	\onecolumngrid
	\newpage
	\begin{center}
		\large
		
		\textbf{Supplemental Material}
	\end{center}
	\normalsize
	\normalsize
	\twocolumngrid
	
	\section*{Experimental details}
	
	We operate the drive laser at $\SI{790.019}{\nano\meter}$, i.e., the  $^{87}$Rb D-line tune-out wavelength ~\cite{schmidt_precision_2016}, where the scalar ac-Stark shift vanishes for $F=1$. This suppresses dipole potentials and minimizes spontaneous emission, otherwise relevant for the operating powers $\sim\si{100~\mW}$. 
	We experimentally verify that the drive does not induce significant losses by monitoring the atom-number evolution while illuminating with the maximum experimentally used laser power.
	The measured $1/e$ lifetime of \SI{47(7)}{\ms} is orders of magnitude larger than our relevant experimental timescales of $\sim\SI{100}{\micro\s}$.
	Note that during this measurement we keep the cavity unlocked to avoid cavity-induced interactions.  The drive polarization is chosen along $y$ to inhibit atomic self-organization in the $m=\pm1$ sublevels mediated by the vectorial polarizability~\cite{Landini_2018}.
	
	Our optical cavity is a quasi-planar symmetric Fabry-Perot resonator with a length of $\SI{176}{\micro\meter}$ and finesse $\mathcal{F}=3.5\cdot 10^5$; the TEM$_{00}$ mode has a waist of $w_c=\SI{25}{\micro\meter}$.
	The cavity is actively stabilized by locking the resonance to the frequency of a reference laser with wavelength of $\SI{830}{\nano\meter}$ using a Pound-Drever-Hall technique; the $\SI{830}{\nano\meter}$ laser is referenced to the laser generating the transverse pump beams via locking on a transfer cavity.
	
	In Table~\ref{tab1:ExperimentalParametes}, we list all relevant experimental parameters for the measurements shown in this work.
	For the measurements of Fig.~4(b-c) in the main text,
	we adjust the drive power $P$ and thus $\eta \propto \sqrt{P}$ to keep $\chi_\text{+,fix} = -2\pi\times\SI{0.50(2)}{Hz}$ constant.
	We calibrate $P$ using Kapitza-Dirac diffraction~\cite{Gadway:09}. Following Refs.~\cite{ferri_emerging_2021,rosa-medina_observing_2022}, we tune the drive away from the tune-out wavelength and illuminate the atoms in a standing-wave configuration.
	\begin{table}[h]
		\begin{tabular}{SSSSSS} 
			\toprule
			{Fig.} & {$N~(\times 10^4)$ } & {$t~(\SI{}{\micro\second})$}& {$\omega_\text{z}/2\pi$} & {$\delta_{+}/2\pi$ } & 	{$\chi_+/2\pi$ }  \\ 
			&  & &{(MHz)} &{(MHz)}&  {(Hz)}  \\ \midrule
			{1(c)}& & 65&  7.09(1)&  -20.7(3)&  -0.50(1) \\ \midrule
			{2(a,c)} &4.0(3) & 60&  7.09(1)&  -18.7(3)& -0.43(2) \\ 
			{2(b,d)} &5.3(4) & 62&  1.01(1)&  -18.7(3)& -0.43(2) \\ \midrule
			{3(a)} &7.2(6) & 62&  7.09(1)&  -25.7(3)&  -0.33(1) \\ 
			{3(b)} &7.0(6) & 62&  0.90(1)&  -25.7(1)&  -0.33(1) \\ \midrule
			{4(a)}& 8.1(3) & &  7.09(1)&  -22.7(1)&  -0.21(1) \\ 
			{4(b,c)}& 7.9(3) & 80&  7.09(1)&  &  -0.50(2) \\ 
			\bottomrule
		\end{tabular}
		\caption{\label{tab1:ExperimentalParametes}List of experimental parameters.}
	\end{table}
	
	\section*{Detection and calibration of atom populations}
	
	We measure the momentum distribution by shining a high-intensity imaging beam along $y$ on the atoms after \SI{5.5}-\SI{6.2}{ms} of time-of-flight (TOF) expansion. To spatially resolve atoms in the different $m$ sublevels, we apply a magnetic field gradient along $z$ during TOF (Stern-Gerlach separation). We extract the atom populations in the different spin and momentum states from circular-shaped crops in the absorption images. For that, we correct for short-scale intensity variations in the imaging-beam profile~\cite{fletcher2015connecting}. Such variations mainly originate from diffraction effects on the cavity which acts as a thick slit for the light. Our detection is finally calibrated with a systematic uncertainty of $\sim 15\%$ using the dispersive shift of the cavity resonance in the presence of a $m=0$ BEC~\cite{brennecke_optomechanics_2008}. We additionally verify the absence of significant $m$-dependent effects on atom counting. To do so, we drive an initially polarized BEC in $m=-1$ in a three-level Rabi oscillation and observe a constant total atom number while the populations of the three $m$ states change.
	
	The technical detection noise in our system is well captured by a Gaussian distribution~\cite{gross_atomic_2011} as shown by the fit in the right panel of Fig.~2(c) in the main text, yielding a standard deviation of $\sigma_\text{det}\approx 200$. The distributions are then fit by a convolution of a Bose-Einstein distribution $p_\text{BE}(N_\text{p},\expval{N_\text{p}})$  and a normalized Gaussian $G(N_\text{p}, \sigma_\text{G}, \mu)$
	\begin{align}
		p(N_\text{p},\expval{N_\text{p}}, \sigma_\text{G}, \mu)=p_\text{BE}(N_\text{p},\expval{N_\text{p}})*G(N_\text{p},\sigma_\text{G}, \mu)\,,
		\label{eq:Convolution}
	\end{align}
	where $\expval{N_\text{p}}$ is the mean pair number extracted from the raw distribution and $\sigma_\text{G}$ and $\mu$ remain fit parameters to capture position-dependent noise effects; we find $\sigma_\text{G} \approx \sigma_\text{det}$.
	Assuming uncorrelated physical and  technical fluctuations, we estimate the standard deviation of the pair histograms as $\sigma(N_\text{p})=\sqrt{\sigma_\text{exp}^2(N_\text{p})-\sigma^2_\text{G}}$, where $\sigma_\text{exp}(N_\text{p})$ is the experimentally measured standard deviation.
	
	\section*{Heterodyne detection}
	
	We monitor the photon field leaking out of the cavity by separating the $y$- and $z$-polarizations on a polarizing beamsplitter, and detecting each of them with independent heterodyne setups. The latter is used to produce the data discussed in this work: the cavity light field originating from superradiant Raman scattering (at frequency $\omega_\text{SR}$) is fiber coupled and interfered with a local oscillator laser ($\omega_\text{LO}$). The high detection bandwidth of 250 MS/s allows for an all-digital demodulation of the beat notes over a broad frequency range of $2\pi\times[0,125]$~MHz, facilitating frequency-resolved detection  of superradiant photon pulses associated with both scattering channels.
	
	The complex intra-cavity field $\alpha(t)=X(t)+iY(t)$ is obtained from the quadratures $X(t)$ and $Y(t)$ after digital demodulation at a desired target frequency $\omega_T=\omega_\text{SR}-\omega_\text{LO}$. Then, the corresponding power spectral density, PSD($\omega$)=$|\text{FFT}(\alpha)|^2(\omega)$, is calculated using a fast Fourier-transform of the form $\text{FFT}(\alpha)(\omega)=dt/\sqrt{\tilde{N}}\sum_j \alpha^*(t_j) e^{-i\omega t_j}$, where $t_j$ is the time of the $j^\text{th}$ step and $\tilde{N}$ is the total number of steps in the integration window. The traces are divided in time intervals of $T=\SI{150}{\micro\second}$ with an overlap of $50\%$ between subsequent intervals, with the photon number spectrogram being calculated as as $\tilde{n}_\text{ph}(t, \omega) =\text{PSD}(\omega)/T$.
	
	To compute the average photon number traces $n_\text{ph}(t)$ around the target frequency $\omega_T$, we integrate the photon number spectrogram over a small frequency region of $\Omega = 2\pi \times [-200,+200]~\si{\kilo\hertz}$, and obtain
	\begin{align}
		n_\text{ph}(t)=\sum_\Omega\tilde{n}_\text{ph}(t,\omega).
	\end{align}
	Finally, the total number of photons associated with superradiant scattering presented in Fig. 4(b) is obtained by numerically integrating the photon traces in time
	\begin{align}
		\expval*{N_\text{ph}}=2\kappa\int_0^\infty n_\text{ph}(t) dt,
	\end{align}
	with $\kappa=2\pi\times\SI{1.25}{MHz}$ being the cavity field losses.
	
	\section*{Experimental timescales}
	
	The characteristic timescale to produce pairs, $T_\text{int}=2\pi/(N\text{max}\abs{\chi_\pm})$, is determined by the collective couplings $N\chi_\pm$.
	For typical values of $N\approx6\times10^4$ and $\chi_+\approx-2\pi\times\SI{0.4}{Hz}$, we obtain $T_\text{int}\approx\SI{40}{\micro\second}$. 
	On the other hand, the lifetime of the pairs is limited by the harmonic trapping potential since the paired states with $\pm\hbar k$ are not trap eigenstates.
	We estimate the lifetime as $T_\text{LT}=\text{min}(T_\text{exp},~T_\text{sep})$~\cite{hodgman_solving_2017}, with $T_\text{exp}=2\pi/\text{max}(\omega_{hx},\omega_{hz})$ and $T_\text{sep}=R_\text{TF}/v_\text{rec}$ being the characteristic timescales for the expansion in the harmonic trap and for the separation between the pairs and the zero-momentum BEC, respectively.  Using the trap frequencies $[\omega_{hx},~\omega_{hy},~\omega_{hz}]=2\pi\times[204(3),~34(2),~185(1)]~\si{Hz}$, the recoil velocity $v_\text{rec}=0.0058~\text{m/s}$ and a Thomas-Fermi radius of $R_\text{TF}\approx\SI{5.8}{\micro\meter}$, we obtain $T_\text{exp}\approx\SI{5}{ms}$ and $T_\text{LT}=T_\text{sep}\approx\SI{1}{\milli\second}$.
	The separation of timescales, quantified as $T_\text{LT}/T_\text{int}\approx 25$ and $T_\text{LT}/T_\text{coh}\approx 6.7$, ensures that pairs are produced in well-defined individual momentum modes and remain in such throughout the entire dynamics. For comparison, collisionally induced pairs in metastable Helium BECs exhibit a ratio of $T_\text{LT}/T_\text{int}~=~0.7$~\cite{hodgman_solving_2017}, while for Floquet-engineered systems $T_\text{LT}/T_\text{int}~\lesssim~3$~\cite{clark_collective_2017}. Notably, these experiments operate in the spontaneous and weak collective regimes for pair production, respectively.
	
	\section*{Thermal occupation of the paired modes}
	
	We estimate an upper bound for the thermal occupation of the momentum modes forming pairs and conclude that it is negligible compared to quantum fluctuations ($\mathcal{O}(1)$). In our system, with $N \approx 8\times 10^4$ and a mean trap frequency $\bar{\omega}=(\omega_{hx} \omega_{hy} \omega_{hz})^{1/3}=2\pi\times\SI{109}{Hz}$, we get a critical temperature $T_c\approx~\SI{210}{nK}$, and thus a realistic estimation for the cloud's temperature $T\lesssim~\SI{100}{nK}$ for a condensate fraction of $N_\text{c}/N\gtrsim 0.9$~\cite{brennecke_optomechanics_2008}.  
	Due to a Thomas-Fermi density profile, the momentum-space spread of the initial BEC and the produced modes is $\delta k=2\pi /R_{\text{TF}} \approx 0.12k$, in agreement with our absorption images.
	Taking into account the width $\delta k$ of the paired modes, the probability to thermally occupy each of them is 
	
	\begin{align}
		P=2\int_{k-\delta k}^{k+\delta k}\hspace{-5pt} \int_{-\delta k}^{\delta k}\hspace{-2pt} \int_{k-\delta k}^{k+\delta k} \hspace{-10pt} p(\mathbf {k},T)~\dd{k_x}\dd{k_y}\dd{k_z} \approx\hspace{-3pt} ~2.8\times10 ^{-4}
	\end{align}
	for $T=~\SI{100}{nK}$, where $p(\mathbf{k},T)=\mathcal{N}\left[\exp(\frac{E(\mathbf{k})}{k_BT})+1\right]^{-1}$ is the momentum-space probability distribution for the thermal $^{87}$Rb atoms of mass $M$. Here $E(\mathbf{k})=\hbar^2\mathbf{k}^2/(2M)$ is the kinetic energy associated with the momenta $\hbar\mathbf{k}=\hbar(k_x,k_y,k_z)$, $k_B$ the Boltzmann constant and $\mathcal{N}$ a suitable normalization factor.
	
	In our experiment, we prepare a BEC solely in $m = 0$ as the initial step for the generation of pairs by applying a strong magnetic-field gradient to clean spurious atoms in $m=\pm1$~\cite{Landini_2018}.
	From our detection, we can safely assume that the population $N_{\pm 1}$ of these sublevels is $<3\sigma_\text{det}$, which lies within the $99.7\%$ confidence interval of our detection, and thus a total number of thermal atoms in $m=\pm1$ of $N_{\pm1}^{T}<(N-N_\text{c})/N\times3\sigma_\text{det} =60$.
	The average number of thermal atoms occupying the modes $\ket{\pm k}_{\pm1}$ is then $\expval{N_T}=P N_{\pm1}^{T}< 0.016 \ll \mathcal{O}(1)$.
	
	\section*{Derivation of the many-body Hamiltonian}
	
	The Hamiltonian of a single atom dispersively coupled to a single cavity mode by a running-wave laser drive is
	\begin{align}
		\hat{H}_\text{SP}&=\frac{\hat{p}^2}{2M}-\hbar\omega_z\hat{F}_z +\hbar q\hat{F}^2_z+ \hbar\omega_c \hat{a}^\dagger \hat{a} \nonumber\\
		&- i\frac{\alpha_v}{2F}\left[\mathbf{\hat{E}^{(+)}}\times\mathbf{\hat{E}}^{(-)}\right]\cdot\mathbf{\hat{F}}.
		\label{eq:H_SP}
	\end{align}
	Here, the first term denotes the kinetic energy of the atom, $\omega_z/B=2\pi\times700~\text{kHz}/\text{G}$ and $q/B^2=2\pi\times 72~\text{Hz}/\text{G}^2$ are the linear and quadratic Zeeman splittings, and $\mathbf{\hat{F}}=(\hat{F}_x,\hat{F}_y,\hat{F}_z)^T$ is the spin operator for the $F=1$ manifold.
	The operator $\hat{a}^\dagger$ creates a photon in the $z$-polarized cavity mode of frequency $\omega_c$. 
	Because of our choice of drive wavelength, we consider atom-light interactions mediated only by the vectorial polarizability $\alpha_v$. The cavity mode extends along $x$ and has a field amplitude $E_{0}=403~$V/m per photon~\cite{ferri_emerging_2021}. With $E_d$ indicating the amplitude of the drive with frequency $\omega_d$ propagating along $+z$, the negative part of the total electric field is represented by
	\begin{equation}
		\mathbf{\hat{E}^{(-)}}=\frac{E_{d}}{2}e^{ikz}e^{-i\omega_dt}\mathbf{e}_y + E_{0}\cos(kx)\hat{a}\mathbf{e}_z,
		\label{eqSI:total_electric_field}
	\end{equation}
	with unit vectors $\mathbf{e}_j$, $j\in\{x,y,z\}$.
	A unitary transformation $\hat{U}~=e^{i\hat{H}_\text{rot} t/\hbar}$ with $\hat{H}_\text{rot}=\hbar\omega_d\hat{a}^\dagger \hat{a} - \hbar\omega_z\hat{F}_z$, and a global phase rotation $\hat{a} \rightarrow \hat{a}e^{i\pi/2}$ gives
	\begin{equation}
		\begin{split}
			&\hat{H}_\text{SP} = \frac{\hat{p}^2}{2M}-\hbar\delta_c \hat{a}^\dagger\hat{a}+\hbar q\hat{F}^2_z \\
			+&\frac{\alpha_v E_0 E_d \cos(kx)}{8} \qty[\qty(\hat{a}^\dagger e^{ikz}\hspace{-3pt} -\hspace{-3pt} \hat{a}e^{\text{-}ikz}) \qty(\hat{F}_+ e^{\text{-}i\omega_z t} \hspace{-3pt}+\hspace{-3pt} \hat{F}_- e^{i\omega_zt})],
			\label{H_SP_expanded}
		\end{split}
	\end{equation}
	with a drive-cavity detuning $\delta_c=\omega_d-\omega_c$.
	Note that $\abs{\delta_c}<2\pi\times\SI{10}{MHz}$ is small compared to the frequency of the drive, and thus we assume a common wavenumber $k$ for the drive and the cavity.
	We derive the many-body Hamiltonian using a six-mode expansion in momentum and spin space. The selected modes comprise $\ket{0}_0$, with a single-particle wavefunction $\psi\propto 1$, the four modes $\ket{(\pm) k}_{\pm1}$ occupied by pairs, with $\psi \propto \cos{(kx)}e^{(\pm) ikz}$, and the next-order mode $\{k_x, ~k_z, ~m\} = \{\pm 2k, ~0, ~0\} \equiv \ket{\pm2k}_0$, with $\psi\propto \cos^2(kx)$. The latter participates due to the interaction between pairs in $m=\pm 1$.
	The corresponding spinor field operator is
	\begin{align}
		\hat{\Psi}(\mathbf{x})&=\left(\hat{\Psi}_{+1}(\mathbf{x}),\,\hat{\Psi}_{0}(\mathbf{x}),\,\hat{\Psi}_{-1}(\mathbf{x})\right)^T\nonumber \\
		&= 
		\begin{pmatrix}
			\frac{k}{\sqrt{2}\pi}\cos{(kx)}(e^{ikz}\cp+e^{-ikz}\cpm) \\
			\frac{k}{2\pi}\hat{c}_{0,0} + \frac{\sqrt{2}k}{\sqrt{3}\pi}\cos^2(kx)\cx\\
			\frac{k}{\sqrt{2}\pi}\cos{(kx)}(e^{-ikz}\cm+e^{ikz}\cmp)
		\end{pmatrix},
		\label{Eq:FieldOperatator}
	\end{align}
	where 
	the operators $\hat{c}_{\tilde{k},m}$ follow the main-text definitions.
	
	If the two-photon detunings $\delta_\pm= \delta_c\pm\omega_z=(\omega_d\pm\omega_z)-\omega_c=\omega_\pm - \omega_c$ significantly exceed the decay rate $\kappa$, superradiant Raman scattering from $m=0$ to $m=\pm1$ is suppressed. We adiabatically eliminate the cavity field following the formalism of Ref.~\cite{reiter_effective_2012} and obtain the effective many-body Hamiltonian
	\begin{equation}
		\begin{split}
			\hat{H}&= \hat{H}_{0}+\hat{H}_{+}+\hat{H}_{-},\ \ \text{with} \\ \hat{H}_{0}&=\hbar\frac{\omega_0}{2}(\cp^\dagger\cp+\cpm^\dagger\cpm  \\ 
			&+\cm^\dagger\cm+ \cmp^\dagger\cmp)\\ 
			&+4\hbar\omega_\text{rec}\cx^\dagger\cx, \\
			\hat{H}_{+}&=\hbar\chi_+(2\cm^\dagger\cp^\dagger\czero\czero+\czero^\dagger\cp\cp^{\dagger}\czero \\
			&+\cm^\dagger\czero\czero^\dagger\cm + \text{H.c.}), \\
			\hat{H}_{-}&=\hbar\chi_-(2\cpm^\dagger\cmp^\dagger\czero\czero+\cpm^\dagger\czero\czero^\dagger\cpm \\
			& + \czero^\dagger\cmp\cmp^{\dagger}\czero+\text{H.c.}),  \\
			\label{Eq.H_MB_F}       
		\end{split}
	\end{equation}\\[-0.7cm]
	where we additionally use the notation $\czero~=(\cz+\sqrt{\frac{2}{3}}\cx)$. The four-photon coupling rates,
	given in the main text, depend on the
	two-photon Raman couplings $\eta~=\beta\alpha_vE_0E_d/8\hbar$. The factor $\beta\approx0.89$ arises from the overlap integrals between the harmonically confined atomic cloud, the cavity mode and the drive~\cite{ferri_emerging_2021}. 
	In the $\hat{H}_\pm$ terms, the first part describes the production of pairs in $m=\pm1$ starting from $m=0$, while the second (third) describes spin-exchange interactions between $m=0\leftrightarrow m=1$ ($m=0\leftrightarrow m=-1$). In presenting the Hamiltonian in Eq.~(2) of the main text we omit the latter parts, as spin-exchange dynamics is suppressed when the majority of the atoms populate the pump mode. 
	In the simulations, however, we take into account the full Hamiltonian.
	
	\section*{Cavity Dissipation \\  and Equations of Motion}
	
	The leakage of the cavity field makes our experiment intrinsically an open quantum system. We identify effective Lindblad terms
	\begin{align}
		\hat{L}_{\pm}=&\sqrt{\gamma_\pm}\left(\hat{c}^\dagger_{+k, \pm 1} \czero+\czero^\dagger\hat{c}_{-k, \mp 1}\right), 
		\label{Eq.L_MB}
	\end{align}	
	which we derive within the framework of the effective operator formalism~\cite{reiter_effective_2012}. The term $\hat{L}_{\pm}$  describes  a superradiant Raman decay process, where atoms scatter photons into the cavity while changing their spin state $m\rightarrow m\pm1$ and obtaining net recoil momentum $+\hbar k$ along $z$. These cavity photons get lost before they can be further rescattered. The dynamics of the open quantum system is determined by the master equation
	\begin{align}
		\small
		\frac{d\hat{\rho}}{dt}=-\frac{i}{\hbar}\left[\hat{H},\hat{\rho}\right] \hspace{-2pt}+\hspace{-5pt}\sum_{j\in\{+,-\}}\hspace{-5pt} \hat{L}_j\hat{\rho} \hat{L}_j^\dagger-\frac{1}{2}\left(\hat{L}_j^\dagger\hat{L}_j\hat{\rho} +\hat{\rho} \hat{L}_j^\dagger\hat{L}_j\right).
	\end{align}
	We define complex-valued expectation values $\psi_{\tilde{k},m}=\expval{\hat{c}_{\tilde{k},m}}$ for the different modes and derive mean-field equations of motion (EOM)
	\begin{small}
		\begin{equation}
			\begin{split}
				\frac{d}{dt}\psi_{0,0}&\\
				=-i\chi_+&(2\psi_0^*\psi_{\text{-}k,\text{-}1}\psi_{\text{+}k,\text{+}1} +\psi_{\text{+}k,\text{+}1}^*\psi_0\psi_{\text{+}k,\text{+}1} +\psi_{\text{-}k,\text{-}1}^*\psi_0\psi_{\text{-}k,\text{-}1} )\\
				-i\chi_-&(2\psi_0^*\psi_{\text{+}k,\text{-}1}\psi_{\text{-}k,\text{+}1} +\psi^*_{\text{-}k,\text{+}1}\psi_0\psi_{\text{-}k,\text{+}1} +\psi_{\text{+}k, \text{-}1}\psi_0\psi^*_{\text{+}k, \text{-}1})\\
				+\gamma_+&(\psi^*_{\text{+}k,\text{+}1}\psi_{\text{+}k,\text{+}1}\psi_0 - \psi^*_{\text{-}k,\text{-}1}\psi_{\text{-}k,\text{-}1}\psi_0) \\
				+\gamma_-&(\psi^*_{\text{+}k,\text{-}1}\psi_{\text{+}k,\text{-}1}\psi_0 - \psi^*_{\text{-}k,\text{+}1}\psi_{\text{-}k,\text{+}1}\psi_0)  \\
				\frac{d}{dt}\psi_{\pm k,\pm1}&=-i\frac{\omega_0}{2}\psi_{\pm k,\pm1} \\
				&\pm (\gamma_+\mp i\chi_+)(\psi_0^*\psi_0\psi_{\pm k,\pm1} + \psi_{\mp k,\mp1}^*\psi_0\psi_0 ) \\
				\frac{d}{dt}\psi_{\mp k,\pm1}&=-i\frac{\omega_0}{2}\psi_{\mp k,\pm1}\\
				&\pm (\gamma_-\mp i\chi_-)(\psi_0^*\psi_0\psi_{\mp k,\pm1} + \psi_{\pm k,\mp1}^*\psi_0\psi_0 ) \nonumber
			\end{split}
		\end{equation}
	\end{small}
	
	\begin{small}
		\begin{equation}
			\begin{split}
				\frac{d}{dt}\psi_{\pm 2k_x,0}&=-4i\omega_\text{rec}\psi_{\pm 2k_x,0} \\
				+\sqrt{\frac{2}{3}}\large[-i\chi_+&(2\psi_0^*\psi_{\text{-}k,\text{-}1}\psi_{\text{+}k,\text{+}1}+ \psi_{k,1}^*\psi_0\psi_{\text{+}k,\text{+}1}+\psi_{\text{-}k,\text{-}1}^*\psi_0\psi_{\text{-}k,\text{-}1})\\
				-i\chi_-&(2\psi_0^*\psi_{\text{+}k,\text{-}1}\psi_{\text{-}k,\text{+}1}+\psi^*_{\text{-}k,\text{+}1}\psi_0\psi_{\text{-}k,\text{+}1}+\psi^*_{\text{+}k, \text{-}1}\psi_0\psi_{\text{+}k,\text{-}1}) \\
				+\gamma_+&(\psi_{\text{-}k,\text{-}1}^*\psi_0\psi_{\text{-}k,\text{-}1}-\psi_{\text{+}k,\text{+}1}\psi_0\psi^*_{\text{+}k,\text{+}1}) \\
				+\gamma_-&(\psi_{\text{-}k,\text{+}1}^*\psi_0\psi_{\text{-}k,\text{+}1}-\psi_{\text{+}k,\text{-}1}\psi_0\psi_{\text{+}k,\text{-}1}^* )].
				\label{Eq.EOMS}
			\end{split}
		\end{equation}
	\end{small}
	
	\section*{Truncated Wigner Simulations}
	
	We employ truncated Wigner simulations to model the dynamics of our system in the presence of technical and quantum fluctuations, closely following the methodology proposed for interacting Bose gases in Ref.~\cite{blakie_dynamics_2008}. Within this approximation, the system is truncated to relevant empty excitation modes $\hat{c}_{q}$ that are represented by stochastic complex variables $\psi_q$, with $\hat{c}_{q}\in\{\cp,\ \cm, \ \cpm, \ \cmp, \ \cx \}$ . If the occupation of the different modes is initially uncorrelated, they can be sampled from suitable Gaussian-shaped Wigner distributions with mean $\expval*{\psi_q}=0$, and  standard deviations $\sigma[\text{Re}(\psi_q)]=1/2$ and $\sigma[\text{Im}(\psi_q)]=1/2$. This yields effective initial occupations of  
	\begin{align}
		\expval{N_\text{QF}}=\expval*{\hat{c}_q^\dagger\hat{c}_q}=\expval*{\psi^*_q\psi_q}=1/2
		\label{eq6:quantumOneHalf}
	\end{align}
	for the empty excitation modes, typically referred to as \textit{quantum one-half fluctuations} and interpreted as the degree of  vacuum noise relevant for the dynamics of the system~\cite{mink_variational_2022}.
	
	We initialize all the atoms in the mode $\cz$ by setting $\psi_{0,0}(t=0)=\sqrt{N}$, and sample all the other modes from complex-valued normal distributions with $\mu=0$ and $\sigma[\text{Re}(\psi_q)]=\sigma[\text{Im}(\psi_q)]=1/2$. Practically, we sample $>500$ different initial conditions for the mean-field EOM in Eq.~\eqref{Eq.EOMS}, which we then numerically evolve using built-in MATLAB methods. We also incorporate shot-to-shot fluctuations of the initial atom number, on the order of $\Delta N/N=0.05$. We sample the atom number for each simulation from a Gaussian distribution with mean $N$ and standard deviation $\sigma(N) =0.05N$. Finally, we estimate the expectation value $\expval{\hat{O}(t)}$ and the variance $\sigma^2[\hat{O}(t)]$ of the observable $\hat{O}(t)$ at any given time $t$ by averaging over the different samples.
	
	\section*{Simulated statistics for vacuum-stimulated pair production and classical seeds}
	
	The highly nonlinear amplification of pair production allows to characterize the nature of the initial quantum state by measuring the resulting pair statistics after a quench. For this purpose, we perform truncated Wigner simulations with a variable number of classical pairs $\seed$ seeding the dynamics. Practically, we sample the individual simulations from Gaussian distribution with $\mu[\text{Re}(\psi_q)]=\sqrt{N_\text{p}^\text{s}}$.
	\begin{figure}[h]
		\centering
		\includegraphics[width=\columnwidth]{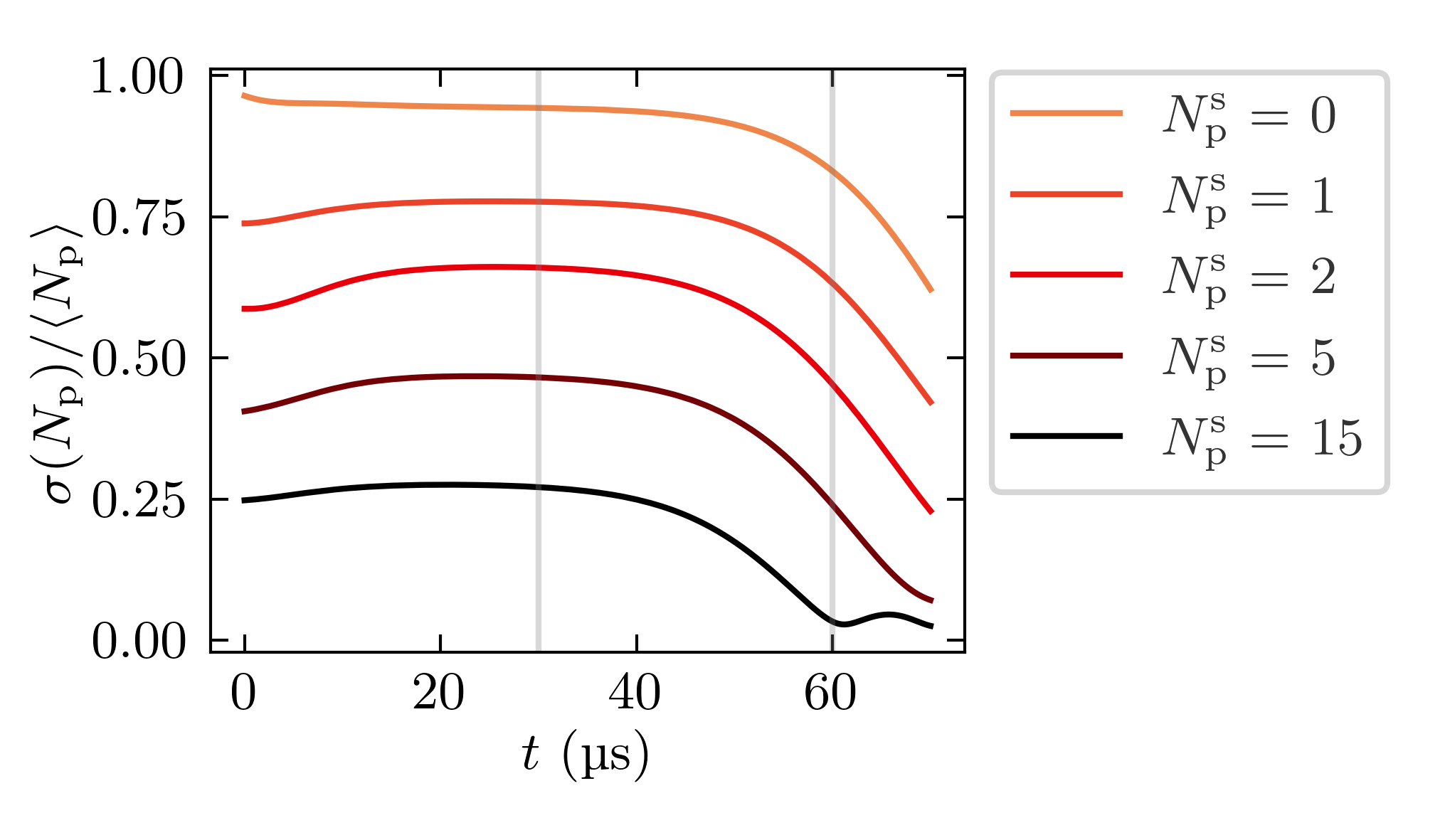}
		\caption{Simulated ratio of standard deviation to mean number of pairs $\sigmapairs/\expval{\pairs}$ for an initial vacuum state and different classical seeds $\seed$ as a function of quench time $t$. The vertical gray lines mark the times of the histograms shown in Fig.~\ref{fig:SI_std_over_mean}. For the parameters of the truncated Wigner simulations, see Tab.~\ref{tab1:ExperimentalParametes} and Fig. 2(c) in the main text.
		}   
		\label{fig:SI_std_over_mean}
	\end{figure}
	\begin{figure*}[t]
		\centering
		\includegraphics[width=1.8\columnwidth]{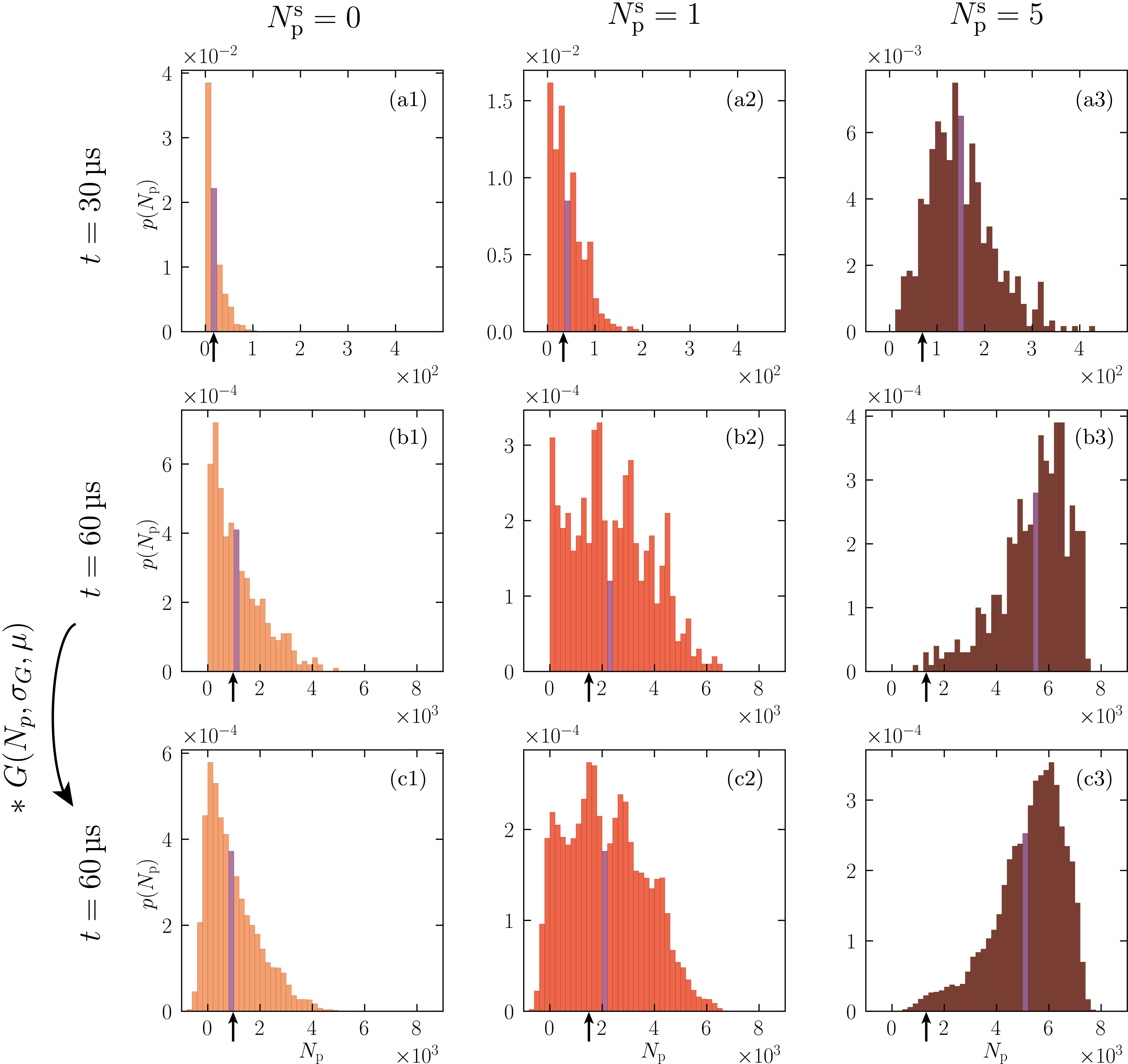}
		\caption{Simulated pair statistics for vacuum and different classical seeds. 
			(a)~Pair histograms for short evolution times ($t=\SI{30}{\micro\second}$) and representative seeds of $\seed=0$ (a1), 1 (a2) and 5 (a3). (b,c) Corresponding histograms for the experimentally relevant evolution time ($t=\SI{60}{\micro\second}$), directly obtained from the simulations (b) and after convolution with the technical detection noise (c).  The mean and standard deviation for each distribution are marked by the purple bin and the black arrows, respectively. For the parameters of the truncated Wigner simulations, see Tab.~\ref{tab1:ExperimentalParametes} and Fig. 2(c) in the main text.
		}   
		\label{fig:SI_histograms}
	\end{figure*}
	In Fig.~\ref{fig:SI_std_over_mean}, we plot the ratio of the standard deviation $\sigmapairs$ and the mean pair number $\expval{\pairs}$ for an initial vacuum state ($\seed=0$) and different classical coherent states ($\seed>0$) as a function of quench time $t$ from 500 simulations. We observe that for an initial vacuum state ($\seed=0$), this ratio is close to unity as expected for a Bose-Einstein distribution. Already a small classical seed comprising a single pair ($\seed=1$) results in a significant reduction of $\sigmapairs/\expval{\pairs}$, as the mean pair number is over-proportionally amplified compared to the quantum fluctuations of the corresponding coherent state; larger seeds further enhance this effect.
	For quench times up to $t\lesssim\SI{40}{\micro\second}$, the undepleted pump approximation is well-justified and the ratio stays approximately flat. At the experimental quench time $t=\SI{60}{\micro\second}$ employed for our measurements in Fig.~2 of the main text, pump depletion becomes non-negligible and results in a reduction of the mean-to-standard deviation ratio, especially for the classically seeded dynamics. This regime is particularly sensitive to small classical seeds, since the resulting pair number distributions are strongly altered.
	
	In Fig.~\ref{fig:SI_histograms}, we present simulated histograms obtained from vacuum fluctuations and different classical seeds for early and experimentally relevant quench times.
	For the short quench time $t=\SI{30}{\micro\second}$, we observe the expected Bose-Einstein distribution for vacuum-seeded dynamics with $\expval{\pairs}\approx\sigmapairs$, see Fig.~\ref{fig:SI_histograms}(a1).
	Even a classical seed of one pair $N_\text{p}^\text{s}=1$ is sufficient to significantly alter the histogram's shape, whereas a seed of $N_\text{p}^\text{s}=5$ additionally shifts the mean significantly, see Figs.~\ref{fig:SI_histograms}(a2-a3).
	For the experimentally relevant quench time $t=\SI{60}{\micro\second}$, these effects become more pronounced, as also the skewness of the distributions progressively changes from positive to negative with increasing seeds, see Fig.~\ref{fig:SI_histograms}(b1-b3). To directly compare these simulations with our experimental results, we convolve the simulated histograms with our Gaussian detection noise $G(N_\text{p}, \sigma_G, \mu)$, see Fig.~\ref{fig:SI_histograms}(c1-c3) and Fig.~2(e) of the main text. Effectively, the convolution (see also Eq.~\eqref{eq:Convolution}) smooths and slightly shifts the distributions while preserving their characteristic shape important for revealing vacuum-seeded dynamics.
	
	\section*{Analogy to spontaneous parametric down-conversion}
	
	In the following, we elaborate on the analogy between our two-channel configuration, spontaneous parametric down-conversion (SPDC) and Bell states, closely following the approach outlined in Refs.~\cite{lewis-swan_proposal_2015, thomas_matter-wave_2022}. The interaction Hamiltonian for the two-channel configuration using the undepleted pump approximation $\hat{c}_{0,0}\approx \sqrt{N}$ reads:
	\begin{align}
		\hat{H}_\text{int} &= \hat{H}_+ + \hat{H}_-\notag \\
		&\approx 2\hbar N \chi_+(\hat{c}^\dagger_{\--k,-1}\hat{c}^\dagger_{+k,+1} +\text{H.c.}) \notag\\ 
		&+2\hbar N \chi_-(\hat{c}^\dagger_{\--k,+1}\hat{c}^\dagger_{+k,-1} +\text{H.c.}),
		\label{eq:H_int_two-channel}
	\end{align}
	where we neglect the spin exchange terms. With the definitions $g_\pm=2N\chi_\pm$, we can rewrite Eq.~\eqref{eq:H_int_two-channel} in the well-known form of non-collinear, type-II SPDC~\cite{pan_multiphoton_2012}
	\begin{align}
		H_\text{SPDC} = \hbar \qty(g_+ \hat{c}^\dagger_{-k,-1}\hat{c}^\dagger_{+k,+1} +  g_-\hat{c}^\dagger_{\--k,+1}\hat{c}^\dagger_{+k,-1}) +\text{H.c.},
		\label{eq:H_SPDC}
	\end{align}
	where the spin states $m=\pm 1$ correspond to horizontally and vertically polarized photons propagating with well-defined momenta $\pm k$ along two distinct spatial directions ($\pm z$).
	We assume an initial vacuum state 
	\begin{align}
		\ket{\psi}_0=\ket{0}_{+k,+1} \ket{0}_{-k,-1} \otimes \ket{0}_{+k,-1} \ket{0}_{-k,+1},
		\label{eq:vacuum_state}
	\end{align}
	with $\ket{\pairs}_{\tilde{k},m}$ being a Fock state in the corresponding pair modes. When quenching the couplings $g_\pm$ for a time $t$, the Hamiltonian of Eq.~\eqref{eq:H_SPDC} generates a product state of two-mode squeezed vacuum states~\cite{braunstein_quantum_2005}
	\begin{align}
		\ket{\psi} &= \ket{\psi_+} \otimes \ket{\psi_-},
		\label{eq:product_state}
	\end{align}
	with $\ket{\psi}_+=\sum^\infty_{N_\text{p}=0}\sqrt{p_\text{BE}(N_\text{p})}\ket{N_\text{p}}_{+k, +1}\ket{N_\text{p}}_{-k, -1}$, and $\ket{\psi_-}$ accordingly for the other channel $\chi_-$. Using the mean pair number for each channel $\expval{N_\text{p}^\pm}=\sinh^2(g_\pm t)$~\cite{linnemann_quantum-enhanced_2016, qu_probing_2020}, the coefficients of the different twin-Fock states contributing to $\ket{\psi_\pm}$ can be restated as $\sqrt{p_\text{BE}(N_\text{p}^\pm)}=\sqrt{1-\tanh^2(g_\pm t)}\tanh^{N_\text{p}}(g_\pm t)$. Defining $\mu_\pm = \tanh(g_\pm t)$, we rewrite the two-mode squeezed vacuum states as
	\begin{align}
		\ket{\psi_+} &= \sqrt{1-\mu_+^2}\sum_{N_\text{p}^+=0}^\infty \mu_+^{N_\text{p}^+} \ket{N_\text{p}^+}_{+k, +1}\ket{N_\text{p}^+}_{-k, -1}, \notag \\
		\ket{\psi_-} &= \sqrt{1-\mu_-^2}\sum_{N_\text{p}^-=0}^\infty \mu_-^{N_\text{p}^-} \ket{N_\text{p}^-}_{+k, -1}\ket{N_\text{p}^-}_{-k, +1}.
	\end{align}
	For clarity, we consider first the case of small Zeeman splittings $\omega_z\to 0$, where both couplings $\mu_+\approx \mu_-=\mu$ become equal. In this limit, we can explicitly write Eq.~\eqref{eq:product_state} as 
	\begin{align}
		\ket{\psi} = (1-\mu^2)\sum_{N_\text{p}^+, N_\text{p}^-=0}^\infty \mu^{N_\text{p}^+ + N_\text{p}^-} &\ket{N_\text{p}^+,N_\text{p}^+;N_\text{p}^-,N_\text{p}^-},
		\label{eq:double_TMSV_same_coupling}
	\end{align}
	where for conciseness we omit the labels for spin and momentum and identify the modes via their positions in Eq.~\eqref{eq:vacuum_state}.
	In the high gain-limit $\mu\to 1$, as considered in this work, the state of Eq.~\eqref{eq:double_TMSV_same_coupling} offers exciting possibilities for future experiments to study bipartite spin entanglement in spatially separated macroscopic atomic clouds along the drive $\pm z$-direction, see also Ref.~\cite{lange_entanglement_2018}.
	
	In the following, we focus on the low-gain limit $\mu\ll 1$, which opens up new avenues for prospective Bell tests with massive particles.
	Here, Eq.~\eqref{eq:double_TMSV_same_coupling} can be truncated to states corresponding to a single pair produced via either channel:
	\begin{align}
		\ket{\psi} \approx (1-\mu^2)\Big(\ket{0,0;0,0} + \mu\ket{1,1;0,0} + \mu\ket{0,0;1,1}\Big).
	\end{align}
	Since the vacuum state $\ket{0,0;0,0}$ does not contribute to any of the measurements considered, we reduce the above state to
	\begin{align}
		\ket{\psi} = \frac{1}{\sqrt{2}}\Big(\ket{1,1;0,0} + \ket{0,0;1,1}\Big),
		\label{eq:Bell_state_spins}
	\end{align}
	The state of Eq.~\eqref{eq:Bell_state_spins} describes a pair of spin-entangled $m=\pm 1$ atoms propagating along $+z$- and $-z$-direction. This state is analogous to a maximally polarization-entangled Bell state of two photons in two distinct spatial modes generated via SPDC~\cite{pan_multiphoton_2012}.
	Explicitly, we can identify the spin states $m=+1$ and $m=-1$ with the two polarization states horizontal $(H)$ and vertical $(V)$, respectively:
	\begin{align}
		\ket{1}_{+1,+k} \ket{0}_{-1,+k} &\longrightarrow \ket{H}_{z} \notag\\ 
		\ket{0}_{+1,-k} \ket{1}_{-1,-k} &\longrightarrow \ket{V}_{-z} \notag\\[2ex]
		\ket{0}_{+1,+k} \ket{1}_{-1,+k} &\longrightarrow \ket{V}_{z} \notag \\
		\ket{1}_{+1,-k} \ket{0}_{-1,-k} &\longrightarrow \ket{H}_{-z}.
		\label{eq:mapping}
	\end{align}
	Therefore, Eq.~\eqref{eq:Bell_state_spins} can be rewritten as a Bell triplet state
	\begin{align}
		\ket{\psi}_\text{Bell}^\text{max} = \frac{1}{\sqrt{2}} \Big(\ket{H}_z\ket{V}_{-z} + \ket{V}_z\ket{H}_{-z}\Big).
		\label{eq:Bell_state_spins_mapped}
	\end{align}
	
	Finally, we consider the case of finite Zeeman splittings $\omega_z\neq 0$, where both couplings become unequal $\mu_+\neq \mu_-$. The truncated state in the low-pair limit can then be written as
	\begin{align}
		\ket{\psi} \sim \Big(\mu_+\ket{1,1;0,0} + \mu_-\ket{0,0;1,1}\Big),
		\label{eq:Bell_state_spins_nonmaximal}
	\end{align}
	up to a normalization factor. The state of Eq.~\eqref{eq:Bell_state_spins_nonmaximal} is analogous to a non-maximally entangled Bell state with tunable degree of entanglement via the couplings $\mu_\pm$:
	\begin{align}
		\ket{\psi}_\text{Bell}^\text{non-max} \sim \Big(\mu_+\ket{H}_z\ket{V}_{-z} + \mu_-\ket{V}_z\ket{H}_{-z}\Big)
	\end{align}
	In SPDC, these states are generated through the rotation of the pump beam's polarization~\cite{white_nonmaximally_1999}. This rotation corresponds to manipulation of the magnetic field $B$ in our system, which controls $\omega_z$ and thereby the coupling ratio $\mu_+/\mu_-$. Non-maximally entangled states play a crucial role in Bell tests, and they are especially important for addressing the detection loophole~\cite{brunner_bell_2014}.
	
\end{document}